\documentclass[12pt]{article}
\usepackage{epsf}
\def\be{\begin{equation}}
\def\ee{\end{equation}}
\def\bea{\begin{eqnarray}}
\def\eea{\end{eqnarray}}

\def\yp{\Upsilon}

\begin{document}
\begin{titlepage}
\begin{center}
{\Large \bf William I. Fine Theoretical Physics Institute \\
University of Minnesota \\}
\end{center}
\vspace{0.2in}
\begin{flushright}
FTPI-MINN-12/23 \\
UMN-TH-3110/12 \\
July 2012 \\
\end{flushright}
\vspace{0.3in}
\begin{center}
{\Large \bf $Z_b(10610)$ and $Z_b(10650)$ decays to bottomonium plus pion
\\}
\vspace{0.2in}
{\bf Xin Li$^a$  and M.B. Voloshin$^{a,b,c}$  \\ }
$^a$School of Physics and Astronomy, University of Minnesota, Minneapolis, MN 55455, USA \\
$^b$William I. Fine Theoretical Physics Institute, University of
Minnesota,\\ Minneapolis, MN 55455, USA \\
$^c$Institute of Theoretical and Experimental Physics, Moscow, 117218, Russia
\\[0.2in]

\end{center}

\vspace{0.2in}

\begin{abstract}
We consider the transitions from the $Z_b(10610)$ and $Z_b(10650)$ resonances to the states of bottomonium with emission of a pion. The $Z_b$ resonances are viewed as `molecular' objects of a large spatial size made of heavy $B^{(*)}$ mesons, while the states of bottomonium are considered to be compact, so that an application of the leading (dipole) term in the QCD multipole expansion is assumed to be justified. In this way we calculate the ratios of the decay rates to the final states $\Upsilon(nS)\, \pi$ with $n=1,2,3$, and the ratio of the decay rates to $h_b(kP)\, \pi$ with $k=1$ and 2. We find our estimates in a reasonable agreement with recent experimental data.
\end{abstract}
\end{titlepage}

The isovector `twin' resonances $Z_b=Z_b(10610)$ and $Z_b'=Z_b(10650)$, recently found by the  Belle Collaboration~\cite{bellez} near the respective thresholds $B^* \bar B$ and $B^* \bar B ^*$, are naturally interpreted~\cite{bgmmv,mvw,cghm,mp} as molecular states made of the heavy meson-antimeson pairs.
Each of the two new resonances is  observed through the decay to either $\Upsilon(nS) \, \pi$ with $n=1,2,$ or 3, or to $h_b(kP) \, \pi$ with $k = 1$ or 2. Moreover, the decays to the states of ortho- and para- bottomonium are found to have comparable strength with no suppression of either of them  by the heavy quark spin symmetry.   This behavior is natural within the interpretation of the $Z_b$ resonances as being molecular $S$-wave states of the heavy mesons: $Z_b \sim B^* \bar B - \bar B^* B$ and $Z_b' \sim B^* \bar B^*$, since the total spin of the $b \bar b$ quark pair within a meson system is not fixed~\cite{bgmmv}. Although at present the type of the threshold singularity in the heavy meson - antimeson channel (bound, virtual, or resonant state) corresponding to the $Z_b$ peaks is not known, it appears clear that the $Z_b$ peaks result from a strong dynamics of very slowly moving mesons near the threshold. The notion that the heavy $b$ and $\bar b$ quarks at the $Z_b$ resonances are moving at distances longer than the characteristic size of bottomonium is also in a qualitative agreement with the recently available data~\cite{bellebr} on the yield 
of different radial excitations of bottomonium in the decays of these resonances to $\Upsilon(nS) \, \pi$ and $h_b(kP) \, \pi$. Namely, the yield does not diminish with the number of excitation in spite of kinematical suppression for production of heavier states. This implies, at a qualitative level, that the overlap of the bottomonium states with a widely separated heavy quark pair in the initial state increases with the excitation number due to larger spatial size of the excited states. 

The purpose of this paper is to suggest an approach to calculating the relative rates of decay of the $Z_b$ resonances to various radial excitations of bottomonium with emission of light mesons, and thus to quantify the theoretical estimates of the relative strength of the observed transitions $Z_b^{(')} \to \Upsilon(nS) \, \pi$ and $Z_b^{(')} \to h_b(kP) \, \pi$. 
We assume that the bottomonium $b \bar b$ system in the final state is pure color singlet and is sufficiently compact, so that its interaction with soft gluon field can be considered within the multipole expansion in QCD~\cite{gottfried,mv78} with the leading term being the chromo-electric dipole. The transition of the heavy $b \bar b$ pair from the initial `molecular' state to bottomonium is due to this interaction at short distances whose scale is set by the bottomonium size, while the (soft) gluon field induces the transition of the light quark-antiquark-gluon components of the initial `molecule' to the light hadron(s) in the final state. In this picture the specific form of the heavy $b \bar b$ `overlap' amplitude is set by the wave function of the initial state, the chromo-electric dipole interaction, and the wave function of the bottomonium state. Therefore, given a model for the latter wave function for various radial excitations, one can evaluate the relative strength of the transitions to those excitated states of bottomonium. In our estimates in this paper we use the simple model with the Cornell potential~\cite{cornell}. As for the initial state wave function of the $b \bar b$ pair we use the short-distance part of that for a slowly moving pair. Moreover, the chromo-electric interaction links the color singlet finite state to a color-octet initial $b \bar b$ pair. Clearly, such state is present in a molecular heavy meson-antimeson system. Indeed, in the colorless $B$ or $B^*$ mesons the color of the $b$ quark is correlated with the color of the light antiquark. Then in a well separated meson-antimeson system the color of $\bar b$ in the meson is fully uncorrelated with that of $b$ in the antimeson, so that a color-octet $b \bar b$ pair is present with the statistical weight 8/9 and the statistical weight of a colorless $b \bar b$ state is 1/9. At short distances there is a weak (Coulomb-like) repulsion between $b$ and $\bar b$ in the color octet state. Although suppressed by the color factor $1/(N_c^2-1)=1/8$ the effect of this repulsion is noticeable at a small momentum of the heavy quarks, and we take it into account.
Furthermore, it is important for the discussed approach that, even though the orbital angular momentum of the heavy mesons in a molecular system can be fixed ($S$ wave in the $Z_b$ resonances), the orbital angular momentum of the $b \bar b$ pair in such system is generally not fixed~\cite{bgmmv,mvw} due to the motion inside the mesons. Thus the chromo-electric dipole transitions to the $S$-wave $\yp(nS)$ of bottomonium proceed from the $P$-wave state of the initial $b \bar b$ pair, while those transitions to the $P$-wave $h_b(kP)$ levels are dominantly from the initial $b \bar b$ $S$-wave pair, since the wave function for a $D$-wave is suppressed at short distances.

In the calculation in this paper we use the Hamiltonian for the chromo-electric dipole interaction in the form
\be
H_{E1}=-{1 \over 2} \, \xi^a \, {\vec r} \cdot {\vec E}^a (0)~,
\label{he1}
\ee
where $\xi^a=t_1^a-t_2^a$ is the difference of the color generators
acting on the quark and antiquark (e.g. $t_1^a = \lambda^a/2$ with
$\lambda^a$ being the Gell-Mann matrices),  ${\vec r}$ is the vector
for relative position of the heavy quark and the antiquark. Finally, ${\vec E}$ is the chromo-electric gluon field strength. We therefore write the amplitudes of the discussed decays in the form
\bea
&&\left \langle \Upsilon(nS) \, \pi | H_{E1} | Z_b \right \rangle = C_S \, A_{nS} \, E_\pi \, (\vec Z  \cdot \vec \Upsilon) \nonumber \\
&&\left \langle h_b(kP) \, \pi | H_{E1} | Z_b \right \rangle = C_P \, A_{kP} \, \left ( \vec p_\pi \cdot [ \vec Z \times \vec h ] \right )~,
\label{asp}
\eea
where $E_\pi$ ($\vec p_\pi$) is the pion energy (momentum), $\vec Z, ~ \vec \Upsilon$, and $\vec h$ are the polarization amplitudes of the initial and final resonances, and $C_S$, $C_P$ are constants that do not depend on the excitation number of the final state, while this dependence is contained in the amplitudes $A_{nS}$ and $A_{kP}$ describing the overlap integrals with the dipole interaction (\ref{he1}):
\be
A_{nS}= \int R_{nS}(r) \, r \, R_P^{(8)}(r) \, r^2 \, dr~,
\label{ans}
\ee
\be
A_{kP}= \int R_{kP}(r) \, r \, R_S^{(8)}(r) \, r^2 \, dr~.
\label{akp}
\ee
In the latter expressions $R_{nS}$ ($R_{kP}$) are the radial wave functions of the bottomonium $S$ ($P$) wave states and $R_S^{(8)}$ ($R_P^{(8)}$) are the radial wave functions of the color-octet $b \bar b$ pair at small momentum above the threshold in the corresponding partial wave.

It can be noted that the constants $C_S$ and $C_P$ encode the information about the amplitudes for the $b \bar b$ pair to be in the corresponding color and orbital state as well as the amplitude for the conversion by the gluon operator $\vec E$ of the initial light quark-antiquark-gluon `environment' into the final pion.  Clearly, these constants are beyond present theoretical control, and for this reason it is not possible within the present approach to establish a quantitative relation between transitions to $S$- and $P$-wave states of bottomonium. The only guidance on the behavior of the light-hadron part of the amplitudes in Eq.(\ref{asp}) is provided by the soft-pion properties, which mandate the factor $E_\pi$ in the transitions to $\Upsilon(nS)$ and the factor $\vec p_\pi$ in the transitions to $h_b(kP)$~\cite{bgmmv}. Once these factors are accounted for as in Eq.(\ref{asp}),  all the dependence on the excitation number of the specific final bottomonium state is contained in the overlap integrals (\ref{ans}) and (\ref{akp}).

In order to evaluate the latter overlap integrals we use the potential model of heavy quarkonium with the Cornell potential~\cite{cornell}
\be
V=-{\kappa \over r} + {r \over a^2}
\label{vc0}
\ee
with $\kappa = 0.52$ and $a=2.34\,$GeV$^{-1}$, and calculate numerically the eigenfunctions $R_{nS}$ and $R_{kP}$ (we also set $m_b=5\,$GeV). We further consider the relative momentum $q$ of the $b$ and $\bar b$ quarks in the initial state as small. In the limit, where the Coulomb-like repulsion in the octet state is neglected, in the limit of small $q$ the radial function in the $S$-wave state can be considered as constant $R_S^{(8)}(r) \approx \,$const, while that in the $P$ can be set as proportional to $r$: $R_P^{(8)}(r) \approx {\rm const} \, r$. (Clearly, the overall normalization of these functions is not important for calculation of the ratios of the amplitudes $A_{nS}$ with different $n$ and the ratios of $A_{kP}$ with $k=1$ and 2.) In what follows we consider the modification of the overlap integrals in Eqs.~(\ref{ans}) and (\ref{akp}) by the short-dsitance effect of the Coulomb-like repulsion between $b$ and $\bar b$ in the color octet state. This repulsion is described by the potential 
\be
V_8(r)= {\kappa_8 \over r}~,
\label{vc8}
\ee
and  we use in our estimates the value of 
the coefficient $\kappa_8$  related to that in the color singlet (Eq.(\ref{vc0})) as in a one gluon exchange: $\kappa_8= \kappa/8 = 0.065$.

One point related to the Coulomb-like repulsion that can be mentioned is that if the potential (\ref{vc8}) was applicable at all distances then the continuum wave functions would vanish in the limit $q \to 0$ at any finite $r$ due to impenetrability (from long distances) of the Coulomb barrier at zero energy. However neither the expression (\ref{vc8}) is applicable at long distances, nor the momentum $q$ is set to be literally zero. At longer distances the motion in the molecular state is described by that of heavy mesons, rather than individual heavy quarks, and also the typical values of $q$ in the considered problem are small but finite and are set by the inverse size of the molecule. The overlap integrals are determined by the behavior of the wave function of the $b \bar b$ quark pair at short distances, i.e. at the characteristic size of bottomonium. At these distances the $r$ dependence of the small-momentum wave function can still be calculated in the potential (\ref{vc8}), while the normalization of the wave function is determined by the long-range modification of the interaction 
(\ref{vc8}). Since the normalization of the functions cancels in the discussed here ratios of the amplitudes, one can use in a calculation of the integrals in Eqs.~(\ref{ans}) and (\ref{akp}) either 
the small momentum limit of the continuum wave functions in terms of their short-distance expansion:
\be
R_S^{(8)} = 1 + {\kappa_8 \, m_b \over 2} \, r + O(r^2)~,~~~~ R_P^{(8)} = r \, \left [ 1 + {\kappa_8 \, m_b \over 4} \, r + O(r^2) \right ]  
\label{csd}
\ee
(these terms do not depend on $q$), or introduce a small but finite $q$, and  use the exact Coulomb functions (see e.g. in the textbook~\cite{ll})
\be
R_S^{(8)}= {\rm const} \, e^{i q r} \, _1F_1\left ( 1+i {m_b \, \kappa_8 \over 2 q}; 2; - 2i \, q \, r \right )~, ~~~~
R_P^{(8)}= {\rm const} \, r \,  e^{i q r} \, _1F_1\left ( 2+i {m_b \, \kappa_8 \over 2 q}; 4; - 2i \, q \, r \right )
\label{cwf}
\ee
with $_1F_1(a;b;z)$ being the Kummer confluent hypergeometric function. We apply both approaches and find that they result in similar estimates of the ratios of the considered overlap amplitudes. In particular we find that these ratios only weakly depend on $q$ at $q < 200 \,$MeV, as is expected as long as $q$ is not much larger than $m_b \, \kappa_8/2 \approx 160\,$MeV.

\begin{table}[ht]
\caption{Ratios of decay rates for $Z_b(10610)$} 
\centering 
\begin{tabular}{| c | c c c c |} 
\hline 
Ratio & $\kappa_8=0$ & Eqs.~(\ref{csd}) & Eqs.~(\ref{cwf}), $q=0 \div 0.2\,$GeV  & Experiment~\cite{bellebr} \\ [0.5ex] 
\hline 
${\Gamma[Z_b \to \yp(1S) \pi] \over \Gamma[Z_b \to \yp(2S) \pi]}$ & 0.11 & 0.09 & $0.10 \div 0.11$ & $0.073 \pm 0.029$ \\ [0.5ex]
\hline
${\Gamma[Z_b \to \yp(3S) \pi] \over \Gamma[Z_b \to \yp(2S) \pi]}$ & 0.62 & 0.74 & $0.70 \div 0.60$ & $0.49 \pm 0.19$ \\ [0.5ex]
\hline
${\Gamma[Z_b \to h_b(2P) \pi] \over \Gamma[Z_b \to h_b(1P) \pi]}$ & 0.58 & 0.78 & $0.72 \div 0.63$ & $1.54 \pm 0.95$ \\  [1ex] 
\hline 
\end{tabular}
\label{table:zb1r} 
\end{table}

\begin{table}[ht]
\caption{Ratios of decay rates for $Z_b(10650)$} 
\centering 
\begin{tabular}{ | c | c c c c |} 
\hline 
Ratio & $\kappa_8=0$ & Eqs.~(\ref{csd}) & Eqs.~(\ref{cwf}), $q=0 \div 0.2\,$GeV  & Experiment~\cite{bellebr} \\ [0.5ex] 
\hline 
${\Gamma[Z_b' \to \yp(1S) \pi] \over \Gamma[Z_b' \to \yp(2S) \pi]}$ & 0.10 & 0.08 & $0.09 \div 0.10$ & $0.10 \pm 0.04$ \\ [0.5ex]
\hline
${\Gamma[Z_b' \to \yp(3S) \pi] \over \Gamma[Z_b' \to \yp(2S) \pi]}$ & 0.86 & 1.02 & $0.97 \div 0.83$ & $0.68 \pm 0.24$ \\ [0.5ex]
\hline
${\Gamma[Z_b' \to h_b(2P) \pi] \over \Gamma[Z_b' \to h_b(1P) \pi]}$ & 0.73 & 0.99 & $0.91 \div 0.80$ & $1.99 \pm 1.11$ \\  [1ex] 
\hline 
\end{tabular}
\label{table:zb2r} 
\end{table}

The numerical results of our calculation and the experimental data are presented in the Tables 1 and 2. One can readily see that our estimates are within the range allowed by the current data. It is clear however, that there is much room for improvement of the data as well as for refinement of the theoretical approach. In particular, on the theoretical side, the specific numbers are fully dependent on the model wave functions for the bottomonium states and on a general picture of the motion in a near threshold `molecule'. Once more precise data might become available this may contribute to a better understanding of the structure of both bottomonium and of the molecular states of heavy mesons. Also, in our estimates we used a soft-pion approximation, and ignored any effects of a possible form factor depending on the momentum of the pion, which effects can be especially significant in the transitions to the final state $\yp(1S) \pi$. Such effects may arise from the unknown at present amplitude of the conversion of the light component of the meson-antimeson pair to pion (in the factors $C_S$ and $C_P$) as well as from the recoil factors in the dipole matrix elements $A_{nS}$ and $A_{kP}$. The recoil factor is in fact determined by the process of conversion, namely by the fraction of the pion momentum transferred to individual heavy quark or antiquark as opposed to the recoil against the pair $b \bar b$ as a whole. (This is different from e.g. a photon emission, where the entire photon momentum is transferred to an individual quark or antiquark.) 
We are not aware at present of a proper way of including and estimating these momentum-dependent factors, and for this reason we chose to neglect these altogether. The fact that our numerical result for the relative yield of $\yp(1S) \pi$ is in a reasonable agreement with the data appears to indicate that the form factor effect should not be dramatic.  One can also readily notice that our estimates for the yield of $h_b(2P)$ relative to that of $h_b(1P)$ are about twice smaller than the central values of the experimental data and the agreement is only due to the currently large experimental uncertainty. If future more precise data would change this to a meaningful disagreement, this could possibly point to a considerable deviation from the simple potential model used in our estimates, or indicate an enhanced contribution of dipole transitions to the $P$-wave bottomonium from a continuum $D$-wave of the color-octet $b \bar b$ pair.

MBV acknowledges numerous enlightening discussions of experimental and theoretical issues with Alexander Bondar. The work of MBV is supported, in part, by the DOE grant DE-FG02-94ER40823.

\end{document}